
\def\cm{{\rm\thinspace cm}}
\def\erg{{\rm\thinspace erg}}

\def\g{{\rm\thinspace g}}

\def\km{{\rm\thinspace km}}

\def\Mpc{{\rm\thinspace Mpc}}
\def\Msun{\hbox{$\rm\thinspace M_{\odot}$}}

\def\s{{\rm\thinspace s}}

\def\ergpcmps{\hbox{$\erg\cm^{-3}\s^{-1}\,$}}
\def\ergpcmsqps{\hbox{$\erg\cm^{-2}\s^{-1}\,$}}

\def\ergps{\hbox{$\erg\s^{-1}\,$}}
\def\gpcm{\hbox{$\g\cm^{-3}\,$}}

\def\kmps{\hbox{$\km\s^{-1}\,$}}

\def\kmpspMpc{\hbox{$\kmps\Mpc^{-1}$}}

\def\ref{\par \noindent \hangindent=3pc \hangafter=1}
\def\spose#1{\hbox to 0pt{#1\hss}}
\def\approxlt{\mathrel{\spose{\lower 3pt\hbox{$\sim$}}
	\raise 2.0pt\hbox{$<$}}}
\def\approxgt{\mathrel{\spose{\lower 3pt\hbox{$\sim$}}
	\raise 2.0pt\hbox{$>$}}}
\mathchardef\twiddle="2218

\def\multleft#1{\hbox to size{\vbox {\halign {\lft{##}\cr #1}}\hfill}\par}
\def\multright#1{\hbox to size{\vbox {\halign {\rt{##}\cr #1}}\hfill}\par}

\font\big=cmr10 scaled\magstep2
\font\bigbf=cmbx10 scaled\magstep2

\tolerance=10000
\magnification=\magstep1
\baselineskip=0.7truecm
\hsize=6.2truein
\vsize=9.0truein
\hoffset=0truein
\voffset=0truein
\parindent=0.5truein

\noindent {\bigbf On the influence of X-ray galaxy clusters in the fluctuations
of the Cosmic Microwave Background}

\vskip 2truecm

\noindent {\big M.T. Ceballos \& X. Barcons}

\vskip 1.5truecm

\noindent  {\it Instituto de Estudios Avanzados en F\'\i sica Moderna y
Biolog\'\i a Molecular,
Consejo Superior de Investigaciones Cient\'\i ficas-Universidad de Cantabria,
Avda. Los Castros s/n, 39005 Santander, Spain}

\vskip 1.5truecm

\leftskip 2.5truecm

\noindent {\bf ABSTRACT}

\noindent The negative evolution found in X--ray clusters of galaxies limits
the amount of available hot gas for the inverse Compton scattering of the
Cosmic Microwave Background (the Sunyaev--Zel'dovich effect).  Using a
parametrisation of the X-ray luminosity function  and its evolution in terms of
a coalescence model (as presented in the analysis of a flux limited X-ray
cluster sample by Edge et al. 1990), as well as a simple virialised structure
for the clusters (which requires a gas to total mass fraction $\approxgt 0.1$
in order to reproduce observed properties of nearby clusters) we show that the
Compton distortion $y$ parameter is about two orders of magnitude below the
current FIRAS upper limits.  Concerning the anisotropies imprinted on arcmin
scales they are dominated by the hottest undetected objects. We show that they
are negligible (${\Delta T\over T}\approxlt 10^{-7}$) at wavelengths
$\lambda\approxgt 1$~mm.  At shorter wavelengths they become more important
(${\Delta T\over T}\sim 10^{-6}$ at $\lambda\sim 0.3$~mm), but in fact most
clusters will produce an isolated and detectable feature in sky maps. After
removal of these signals, the fluctuations imprinted by the remaining clusters
on the residual radiation are still much smaller. The conclusion is that X-ray
clusters can be ignored as sources of Cosmic Microwave Background fluctuations.

\medskip

\noindent {\bf Key words:} cosmic microwave background --- galaxies:clustering.

\leftskip 0cm

\vfill\eject

\noindent{\bf 1 INTRODUCTION}

\noindent Clusters of galaxies are the largest gravitationally bound structures
in the Universe.  Their potential wells, possibly created by still undetected
dark matter, contain large amounts of X--ray emitting gas at temperatures of
tens of millions of degrees or more. The amount of gas seen is comparable to
the mass contained in the member galaxies, but it is doubtful that more than
half of the mass of the cluster is contained in these components even when the
gas profile is extrapolated to large distances (see, e.g., Briel et al. 1992;
Mushotzky 1993).

Since energetic electrons are present in the intracluster medium, they inverse
Compton scatter any long wavelength radiation from the background passing
through the Cluster (Sunyaev \& Zel'dovich 1972).  This is particularly
important for the Cosmic Microwave Background (CMB) radiation which constitutes
an overall background source for all clusters. Clusters are optically thin to
the CMB (typically the number of scatterings suffered by a microwave photon in
a cluster is $\tau < 10^{-3}$), and since their covering factor of the sky is
also small, the integrated effect is tiny. Nevertheless recent  upper limits on
the distortion of the CMB spectrum obtained by the FIRAS instrument on board of
COBE (Mather et al, 1993) are indeed very stringent and might challenge models
for cluster evolution (Markevitch et al, 1991).

Furthermore, in  particular lines of sight where clusters with large amounts of
gas are present, the effect might be detectable.  This has been the goal of all
searches for the Sunyaev-Zel'dovich (SZ) effect, using the position of  known
bright X--ray clusters. The effect consists of a net upscattering of the
photons where the average relative change in the frequency of the incoming
photon is ${\Delta\nu\over\nu}\sim 4{kT_{gas}\over mc^2}$ (Rybicky \& Lightman,
1979).  In fact this results in an approximately constant  {\it decrease} of
the brightness temperature in the Rayleigh--Jeans region of the spectrum and a
very steep increase of the temperature at frequencies beyond the CMB blackbody
peak. The Sunyaev-Zel'dovich effect there is  positive and steeply increasing
with frequency.  The extent in frequency of this effect at submillimetre
wavelengths depends on the temperature of the cluster gas upscattering the
background radiation.

A recent result on the X--ray properties of clusters of galaxies is that their
X--ray luminosity function evolves negatively (Edge et al. 1990, Gioia et al.
1990a).  That means that luminous clusters are underrepresented at high
redshifts in comparison with what happens at low redshifts. Quantifying this
effect is not an easy task since it is very much dependent on the sample
selection criteria.  On the one hand, Edge et al. (1990) take a flux limited
all-sky sample ($S(2-10{\rm keV})>1.7 \times 10^{-11}\ergpcmsqps$, with a
redshift depth $z\sim 0.1$) and find different evolution for clusters above and
below a luminosity $\sim 8\times 10^{44}\ergps$. On the other hand, Gioia et
al. (1990a) and Henry et al. (1992)  used the {\it Einstein Observatory}
Extended Medium Sensitivity Survey (EMSS; Gioia et al., 1990b) to find also a
negative evolution in the cluster X-ray luminosity function.  All clusters in
the EMSS sample are less luminous than $8\times 10^{44}\, \ergpcmsqps$ and they
span a much larger redshift range.  In fact, Gioia et al (1990a) find a
significant change in the X-ray luminosity function at $z\sim 0.3$. Therefore
both studies support the idea of negative evolution although in different
luminosity and redshift parts of parameter space (see discussion on these
issues by Henry 1992). However both studies suggest that X--ray luminous
clusters form by merging of smaller sub-clusters.  There is in fact some
evidence for this in specific clusters both from the {\it Einstein Observatory}
(Forman et al. 1981; Gioia et al. 1982; Jones \& Forman 1992) and from {\it
Rosat} in particular in Coma (White, Briel \& Henry 1993, Davis \& Mushotzky,
1993, Briel, Henry \& B\"ohringer 1992), Perseus (Schwarz et al. 1992) and
A2256
(Briel et al. 1991, Davis \& Mushotzky, 1993).  Since the X--ray volume
emissivity is proportional to the square of the gas density, that means that
the amount of gas available for the SZ effect in a typical line of sight is
relatively small.

In this paper, we adopt a parametrisation of the X--ray luminosity function
that correctly describes the flux-limited X-ray cluster sample considered by
Edge et al. (1990) and a simple cluster model where the gas density falls from
a constant within the core radius  to an $r^{-2}$ profile up to a maximum
radius. We will show that this specific cluster model provides a fairly good
description of nearby cluster properties if the gas to total mass ratio is
larger than in the standard Cosmology , i.e., we require ${\Omega_{gas}\over
\Omega_0}\sim 0.15$ when the standard nucleosynthesis value for a flat
Universe is 0.06 ($H_0=50\, {\rm km}\, {\rm s}^{-1}\, {\rm Mpc}^{-1}$ will be
used throughout the paper). We then compute the induced  $y$ Compton distortion
parameter as well as the temperature  fluctuations imprinted in the CMB on
scales of arcmin at different wavelengths.   Our computation indeed assumes
that the coalescence model adopted by Edge et al. (1990) can be extrapolated up
to higher redshifts. We, however, explore different values for the evolutionary
parameter of this model, and the results remain unchanged within a factor of 2
or 3.  There are other works, complementary to this one, where the CMB
temperature fluctuations introduced by the SZ effects are predicted for
specific
theoretical models of large--scale structure of the Universe (Markevitch et al,
1991, 1992,  Cole \& Kaiser 1988, Makino \& Suto 1993, Bartlett \& Silk 1993).
Here we rely uniquely upon observational X--ray data of clusters and the
simplest assumptions and extrapolations to make the appropriate predictions.

In Section 2 we discuss the way in which the clusters are modelled.  Section 3
presents our main computations of the SZ effect at different wavelengths.  We
fully explore the signal produced by these clusters over wavelengths
$\lambda\approxgt 0.3$~mm (where the contribution of the Galaxy is not expected
to be dominant), showing that the hottest ones will be easily detectable at
submillimetre wavelengths and therefore that the statistical fluctuations
induced in the residual CMB will be negligible at all wavelengths.

\medskip

\noindent{\bf 2 THE X--RAY CLUSTER POPULATION}

\noindent {\bf 2.1 The Cluster Model}

\smallskip

\noindent In the simple model we adopted, clusters are considered isothermal
spheres with an electron density profile:

$$ n_e(r) = \cases{n_{e0}, & \quad$r\le r_c$\cr
		    n_{e0}\,{r_c^2\over r^2}, &\quad$r_c\le r \le R$\cr} \eqno
(1)$$

\medskip

\noindent where $n_{e0}$ is the electron number density at the center of the
cluster, $r_c$ is the core radius and $R$ is the total radius of the cluster
out to which gas is present.

Assuming that the clusters are virialized to a radius $R_{vir}$ and that the
gas is in hydrostatic equilibrium, from the spherical infall model (Peebles
1980) we can estimate the gas temperature and the virial radius

$$ T_{gas}= 2.6\times 10^8 \left(M\over 10^{15}\,h_{50}^{-1}\,\Msun\right)
{\mu\over R_{vir}(\Mpc) \,h_{50}} {\rm K}. \eqno (2)$$

$$R_{vir}= {2.71\over (1+z)}\left( M\over
10^{15}\,h_{50}^{-1}\,\Msun\right)^{1/3}h_{50}^{-1} \Mpc. \eqno(3)$$

\medskip

\noindent where $m_p$ is the proton mass, $h_{50}=H_0/50$ \kmpspMpc, $\mu$
is the mean molecular weight of the gas which is $\mu \approx 0.6$ for
primordial abundances, and $M$ is the mass inside $R_{vir}$.

The ratios $R/R_{vir}$ and $R_{vir}/r_c$ are introduced in order to take into
account the possibility that the gas in the cluster extends beyond the virial
radius. Results will be showed for a wide range of these parameters. In
particular we will consider a cluster radius ranging from 1 to 2 virial radii
and a virial radius going from 5 to 10 core radii, which encompasses most of
the reasonable parameters for clusters.

The central density can be  found by assuming that the electron density is
proportional to the mass density in the cluster:

$$ n_e(r) = \rho (r) {f_{gas}\over  m_p}\left(X+1\over 2\right) \eqno
(4)$$ where  $f_{gas}=\Omega_{gas}/\Omega_0$ is the fraction of gas to total
mass in the cluster and $\Omega_0$ will be taken as 1 for simplicity.
$\rho$ is the (total) mass density profile in the cluster and the gas is
considered a fully ionized
mixture of hydrogen and helium with a mass fraction $X=0.76$ of hydrogen. The
above expression indeed assumes that the gas to mass ratio $f_{gas}$ is
constant
independent of environment (see below).


The total cluster mass is,

$$ M = 4\,\pi\,r_c^2\rho_0\left(R-{2\,r_c\over 3}\right) \eqno(5)$$

\noindent whereas the mass within the virial radius is

$$ M_{vir} = 4\,\pi\,r_c^2\rho_0\left(R_{vir}-{2\,r_c\over 3}\right)
\eqno(6)$$

\medskip

These  expressions lead to a relation between total mass and virial mass as
a function of $R/R_{vir}$ and $R_{vir}/r_c$ ratios. Comparing
eq.(6) with the virial mass obtained from eq.(3) the value of the central
density will be given by:

$$ \rho _0 = 2.825 \times 10^{-28}h_{50}^2 \left({R_{vir}\over r_c}\right)^3
\left( {R_{vir}\over r_c}-{2\over 3}\right)^{-1}(1+z)^3 \, \gpcm \eqno(7) $$

With these ingredients we can in principle try to reproduce some of the cluster
properties.  In order to see what are the suitable values for the different
parameters involved (and in particular for $f_{gas}$) we have taken a sample of
25 nearby clusters for which we have luminosities and temperatures (David et
al. 1993) and also the core radius and gas mass within 3 Mpc (Jones \& Forman
1984) and tried to fit the global properties and correlations according to the
model presented here. We have checked that different values of $R/R_{vir}$ do
not produce very significant changes in $f_{gas}$ when we try to describe this
sample according to our model as far as $R/R_{vir}\approxgt 1$. We have
consequently assumed $R=R_{vir}$.

We have considered the Luminosity-Temperature relation
(Figure 1) and the gas mass within 3 Mpc (Figure 2). Fairly good agreement
between expected 2-10~keV luminosity (computed using the measured temperature)
and the observed one is achieved if $f_{gas}\sim 0.15$. In Figure 1 we also
show the luminosity - temperature relation for an average $R_{vir}/r_c=5$   to
10 (at the mean redshift of the sample) and this value of $f_{gas}$. Our
simplified model shows an approximate $L_X\propto T^{\epsilon}$ relation for
constant $f_{gas}$ with $\epsilon\sim 2$. The value found in the whole David et
al. (1993) sample is $\epsilon\sim 3.4$ close to the one found for the EXOSAT
sample by Edge \& Stewart (1991) which is $\epsilon\sim 2.8$. Although this
might imply a change in $f_{gas}$ with mass scale, this point deserves further
study.

A similar effect happens in Figure 2 where we plot the expected versus measured
gas mass within 3 Mpc.  Good agreement is found again for $f_{gas}\sim 0.15$
which is the value adopted here.

In any case the standard Cosmology value $f_{gas}=0.06$ would clearly
underpredict the luminosity for a given temperature as well as the gas mass
within 3~Mpc.  In fact, there is some recent evidence that bright clusters of
galaxies may contain even a larger fraction of baryons to total mass. This is
particularly true in the outskirts of some clusters (Briel, Henry \&
B\"ohringer 1992) where this  ratio may be as large as $\approxgt 0.3$ for
Coma. If we have still underestimated $f_{gas}$ by some significant amount, the
net effect of this in our computations of temperature fluctuations in the CMB
is a {\it decrease} by the same amount.

\medskip

\noindent {\bf 2.2 The distribution of Clusters.}

\smallskip

\noindent Unlike previous work where the distribution of clusters is described
in the framework of the Press-Schechter mass function formalism (Cole \& Kaiser
1988, Makino \& Suto 1993, Bartlett \& Silk 1993), we shall be using a simple
coalescence model used by Edge et al. (1990) to fit their all sky low redshift
cluster distribution.  We shall extrapolate this model (allowing the
evolutionary parameters to take a wide range of values) to the highest
redshifts where clusters are thought to exist ($z\sim 1$).

Detailed work on cosmological models for cluster evolution has been presented
by Kaiser (1991) and Evrard \& Henry (1991).  They both conclude that standard
Cold Dark Matter scenarios with self similar evolution are unable to explain
the shape of the cluster X-ray luminosity function.  In addition different
specific cluster models are built in these works which deviate either from the
expected shape of the power spectrum of the fluctuations or from the
self-similar evolution or from both.  However, since it is not our goal to make
detailed models for the cluster origin and evolution we just take a simple
parametrisation which is consistent with the available data and try to
extrapolate to higher redshifts keeping in mind that the model parameters can
have large uncertainties.

Using a model in which the growth of clusters is due to the merging of
subclusters, Edge et al. (1990) fitted  the evolution of the luminosity
function needed to explain the  detected deficit of luminous clusters at high
redshift. The parametrisation of the X-ray luminosity function proposed can be
written as:

$$\phi(L_x)= {10^A\over B^2}\left({L_x\over 10^{43}}\right)^{-0.6}
\left(1-{1\over B}\right)^{\left[\left({L_x\over
10^{43}}\right)^{0.4}-1\right]}
\quad \Mpc^{-3}{\rm L_{44}^{-1}}.\eqno(8)$$

\medskip

\noindent where $L_x$ is the X-ray luminosity over the 2-10~keV energy band in
$\ergps$, $A=-4.9$, $B=b_1+\left(b_2/(1+z)^{1.5}\right)$ and $L_{44} =
L_x/10^{44} \ergps$. Edge et al. (1990) suggest values $b_1=0.923, b_2=0.82$
However we shall also explore how our
results change when $b_2$ ranges from $0.4$ to $1.1$ keeping $b_1 + b_2 =
1.743$ in order to reproduce de-evolved cluster luminosity function.

There is no evidence whatsoever for a low luminosity cutoff or for a redshift
cutoff, and in fact there is no need for them given the flatness of the
luminosity function and the fact that at a moderate redshift $z\sim 1$ there
are
very few clusters. Thus, in our calculations we extrapolate this function  out
to a redshift $\sim 1$ where we consider all clusters begin to form. However,
we also checked that extending this high redshift cutoff to 2 or 3 does not
produce any change in our conclusions.

With this parametrisation we want to obtain the differential mass function per
unit comoving volume, since all relevant quantities (gas temperature, density
and cluster radius) can be related to the cluster mass and the parameters
$R/R_{vir}$ and $R_{vir}/r_c$ in the simple cluster model presented in the
previous subsection. This requires having the cluster X-ray luminosity in terms
of its mass for a given redshift $z$. In order to do that we recall that to a
good approximation the cluster bremsstrahlung volume emissivity is

$$ \epsilon_{\nu}^{ff}=6.8\times 10^{-38} \sum_i Z_i^2\, n_e\,n_i\,
T_{gas}^{-1/2}\exp\left(-{h\nu\over kT_{gas}}\right)\, \overline
g_{ff}\quad \ergpcmps {\rm Hz^{-1}}\eqno(9)$$

\medskip

\noindent where the sum is extended to all the ion species present (hydrogen
and helium in a fully ionized mixture), $n_e$ and $n_i$ are the electron and
the i-ion number density respectively, $h$ is the Planck constant, $k$ the
Boltzman constant  and $\overline g_{ff}$ is the averaged Gaunt factor which
provides quantum mechanical corrections and that is taken as 1 here. The X-ray
luminosity is then obtained by integrating the above emissivity over the energy
range (2-10~keV in our case) and over the whole cluster volume. The result is

$$\eqalign{&L_x  =3.19\times 10^{45}\,\left({R_{vir}\over r_c}-{2\over
3}\right)^{-2/3}\, \left({R\over r_c}-{2\over 3}\right)^{-4/3}\,\left({4\over
3}-{r_c\over R}\right)\cr &\qquad\left(R_{vir}\over r_c\right)^3
f_{gas}^2\left({X+1\over 2}\right)^2\mu^{1/2}\,h_{50}^{7/3}\,
(1+z)^{7/2}\,m^{4/3}\cr
&\qquad\left[\exp\left(-{0.24\over\mu\,h_{50}^{2/3}\,v^{2/3}\,(1+z)\,m^{2/3}}\right)-
\exp\left(-{1.21\over\mu\,h_{50}^{2/3}\,v^{2/3}\,(1+z)\,m^{2/3}}\right)\right]
\, \ergps\cr}\eqno(10)$$

\smallskip

\noindent where $v={R_{vir}/r_c-2/3\over R/r_c-2/3}$. Since for a given
redshift $z$, this provides a one to one relation between cluster X-ray
luminosity and mass (except for the ratios $R_{vir}/r_c\,$ and $R/R_{vir}\,$
and $f_{gas}$), the number of clusters per unit volume and unit mass
$m=M/10^{15}{\rm M}_{\odot}$ at a redshift $z$ is

$${\cal N}(m;z)=\phi(L_x;z){dL_x(m;z)\over dm}\eqno(11)$$

\medskip

\noindent This expression will be used
to evaluate the Sunyaev-Zel'dovich effect along the line of sight.

As we want to evaluate the SZ effect for the unseen clusters (i.e., those above
the X--ray detection threshold will be avoided in a CMB fluctuation analysis)
we shall take the minimum redshift as $z\sim 0.1$. On the other hand, since we
are going to be interested only in $\sim $ arcmin fluctuations and anisotropies
of the CMB radiation, where the SZ is most relevant, cluster clustering can be
ignored in principle.  At $z=1$ (taken here as
the maximum redshift), one arcmin translates to $0.5\, h_{50}^{-1} \Mpc$.
Since the cluster-cluster spatial correlation function is only fairly well
known at separations larger than a few Mpc (Bahcall \& Soneira 1983, Sutherland
1988), any extrapolation from larger separations might be completely wrong and
therefore we shall adopt the conservative point of view of neglecting it. In
Section 4 we present some evidence that any reasonable clustering amplitude at
the scales we are working does not quantitatively affect our conclusions.

\medskip

\noindent{\bf 3 THE SUNYAEV--ZEL'DOVICH EFFECT}

\noindent {\bf 3.1 The effect produced by a single cluster}

\noindent Cosmic Microwave photons which enter a cluster are upscattered by the
hot electrons in the intracluster gas through a classical Thomson scattering
process since the photon energies are much lower than the electron energies.
The scattered photons will have an isotropic distribution provided that we
consider both the electrons and the incident photons isotropically distributed.

As the mean number of scatterings suffered by the photons is equal to the gas
optical depth, $\tau=\sigma_T \int dl n_e\quad \sim
10^{-2}-10^{-3}$, ($\sigma_T$
being the Thomson cross section and $n_e$ the number density of electrons),
the
inverse Compton scattering process can be evaluated  under the single
scattering approximation. Also since the gas temperature is less than 10 keV
for all the clusters detected so far we will also consider non relativistic
electrons, with a Maxwellian velocity distribution. Under these circumstances
the spectrum of the background radiation along the line of sight observed at
$z=0$ is given by

$$N_{out}(x,z=0) = (1-\tau (z))\,N_{in}(x,z=0)+\tau (z)\int dx'\,
\overline G(x-x';T_{\scriptstyle gas})\,N_{in}(x',z=0)\eqno(12)$$

\medskip

\noindent where $x=\ln\left(h\,\nu\over k\,T_r\right)$, $\nu$ the
frequency of the photon, and
$T_r$ the present temperature of the CMB Radiation which is $2.726\pm 0.01$ K
(Mather et al. 1993).
The Green function $\overline G(x-x';T_{gas})$ gives the probability that a
photon with incident $x'$ is scattered to a value $x$ by the electrons at a
temperature $T_{gas}\,$ (Rybicky \& Lightman, 1979), $\tau(z)$ is the gas
optical depth and $N(x,z)$ is the differential number density of photons at
redshift $z$ (number of photons per unit $x$ per unit volume), i.e.,

$$N(x,z=0,T_b) =8\pi \left({kT_r\over ch}\right)^3 \left( e^{3x}\over
\exp\left(T_r\,e^x\over T_b\right)-1\right)\eqno(13)$$
{}From these changes in the spectrum we may calculate the Sunyaev-Zel'dovich
(SZ)
change in the CMB temperature expressed as:

$${\Delta T\over T}= {T_b-T_r\over T_r}\eqno(14)$$  $T_b$ being the
brightness temperature of the outcoming spectrum.

In Figure~3 we plot this temperature variation for different representative
cases, once the result has been smeared out with  a beam of 1~arcmin FWHM. For
$\lambda > 1$~mm the decrement is negative (the temperature of the radiation
which comes out of the cluster is less than the background incoming
temperature). Nevertheless, in the Wien part of the spectrum ($\lambda< 1$ mm)
there is an important increase of the temperature. This is due in part to the
low value of the intensity of the unperturbed blackbody spectrum at these
wavelengths to be compared with the approximate power-law that the comptonized
spectrum develops in that wavelength domain (see Rybicky \& Lightman 1979).
Therefore the best wavelengths to look for the SZ effect are in principle the
shortest possible ones, as far as the contamination from the galaxy can be
neglected.

\medskip

\noindent {\bf 3.2 The Compton y parameter}

\smallskip

\noindent The Compton distortion parameter defined as  $y=\int
dl\,{k\,T_{gas}\over m_e\,c^2}\,\sigma_{\scriptstyle T}\,n_e$ where $m_e$ is
the electron mass and the integral is evaluated along the line of sight, can be
calculated from the temperature decrement of the background radiation due to
the SZ effect in the Rayleigh--Jeans part of the spectrum ($\lambda> 10$ cm):

$$\left.{\Delta T\over T}\right\vert_{\scriptstyle R-J}=-2\,y
\qquad\hbox{thus}\qquad
<y>=-{<\Delta T>_{\scriptstyle R-J}\over 2\,T_r }\eqno(15)$$

\medskip

The $<\Delta T>_{\scriptstyle R-J}$ shift in the radiation temperature is
evaluated by averaging the effect produced by a single cluster over  the number
density of clusters along the line of sight. We checked that the changes in the
spectrum are quite small and hence they can be related to the change in the
temperature linearly

$$ N(x,T_b) \approx N(x, T_r)\,+ \left.{\partial N\over\partial
T}\right\vert_{T_r}\,(T_b-T_r)\eqno(16)$$

\noindent The average temperature shift can be obtained from the
average change in the photon number density, which is:

$$\eqalign{<&\Delta N(x)> =<N(x)-N_{in}(x)>= \int_{z_{min}}^{z_{max}}
dz\,cH_0^{-1}\,
(1+z)\,(1+\Omega_0z)^{-1/2}\,d_A^2(z)\cr
&\int_{\Omega}d\Omega\,\int_{M_{min}}^{M_{max}} dm\,{\cal N}(m,z)\,\tau(\hat
n,m,z)\, \left[\int\,dx'\,\overline
G(x-x';T_{gas})\,N_{in}(x')-N_{in}(x)\right]\cr} \eqno(17)$$

\medskip

\noindent where $d_A(z)$ is the angular distance (Weinberg 1972), and in
principle we shall use $M_{min} = 10^{13}$ \Msun$\,$ and $M_{max}= 5\times
10^{16}$ \Msun. This last parameter is particularly relevant, since the
flattening of the $\log\,N-\log\,S$ X--ray cluster counts at low fluxes,
reflected in the negative evolution of the luminosity function (Edge et al.
1990), implies that fluctuations will be dominated by the X--ray brightest and,
therefore, more massive objects.

In order to make a proper comparison with current COBE/FIRAS upper limits on
the $y$ parameter we have computed eq. (17) from $z_{min} = 0 $ since in those
observations no extragalactic sources were avoided. Therefore in Figure~4  we
plot $<y>$ as a function of $R/R_{vir}$ for different $R_{vir}/r_c$ ratios.
Although there is a slight increase of the $y$ parameter with $r_c/R_{vir}$
(which is due to the fact that the smaller this parameter is the larger the
central gas density will grow and therefore with less gas the same X-rays are
produced) it is always much smaller than the  COBE/FIRAS present limits, $ y <
2.5\times 10^{-5}$ (Mather et al. 1993).

\medskip

\noindent {\bf 3.3. Induced anisotropies and temperature fluctuations.}

\smallskip

\noindent Since clusters are extended and since the number of clusters in a
typical line of sight is finite (and in fact very small) they imprint
fluctuations and correlations in the CMB maps.  We are now going to compute
these effects by assuming that the space distribution of clusters is uniform at
any given redshift. As mentioned before and discussed with some detail in
Section 4, cluster clustering is very uncertain on scales comparable to the
cluster size ($\sim$~few Mpc). However, even with an extrapolation of the
$r^{-1.8}$ law from larger separations, the effect is negligible on the
results presented here.

If $\theta$ is the angle between two directions represented by the $\hat
n_1,\hat n_2 $ unit vectors, the angular correlation function for the radiation
spectrum is given by

$$\eqalign{C_N(\theta ,x)
&= <(N(\hat n_1,x)-<N(x)>)\, (N(\hat n_2,x)-<N(x)>)>\cr}\eqno(18)$$

\medskip

{}From eq. (18) the angular correlation function is

$$\eqalign{C_N(\theta,x)  &= \int_{z_{min}}^{z_{max}}
dz\,cH_0^{-1}\,(1+z)\,(1+\Omega_0z)^{-1/2}\,d_A^2(z)\cr &\int
d\Omega_{\hat n_1}\,\int_{m_{min}}^{m_{max}} dm\,{\cal
N}(m,z)\,\tau(\hat n_1,m,z)\, \tau(\hat n_2,m,z)\cr &\left[\int\,dx'\,\overline
G(x-x';T_{ gas})\,N_{in}(x')- \,N_{in}(x)\right]^2\cr}\eqno(19)$$

\medskip

Again we use the approximation in eq.~(16) to derive the temperature
correlation function starting from the correlation in the radiation spectrum:

$$C_T(\theta, x)= \left[\left.\left(\partial N(x,T_b)\over \partial
T_b\right)\right\vert_{T_r}\right]^{-2}\, C_N(\theta,x)\eqno(20)$$

\medskip

Due to the finite size of the receiver antenna a smoothing of the intrinsic
correlation function is produced. Thus we approximated the antenna by a
Gaussian of dispersion $\sigma$ in order to compare our calculations and the
results obtained through real experiments. The resulting correlation function
is

$$C_N(\alpha,\sigma,x)={1\over 2\,\sigma}\,\int C_N(\theta,x)\,
exp\left(-{\alpha^2\,+\theta^2\over 4\,\sigma^2}\right)\,
I_0\left(\alpha\,\theta\over 2\,\sigma^2\right)\,\theta\,d\theta\eqno(21)$$

\medskip

\noindent where $C_N(\theta,x)$ is the correlation in the spectrum at a
wavelength $\lambda=h\,c/( k\,T_{r}\,e^x)$, $I_0$ is the modified Bessel
function and $\alpha$ is the beam throw of the telescope. The rms temperature
fluctuations can therefore be evaluated as

$$\left(\Delta T\over T\right)_{rms}={[<(T-<T>)^2>]^{1/2}\over T_r} =
{[C_T(0,\sigma,x)]^{1/2}\over T_r}\eqno(22)$$

\medskip

In Figure~5 we plot the rms fluctuations versus the wavelength for
$R/R_{vir}=1$, $R/R_{vir}=2$ (the extreme ratios) and for two different ratios
$R_{vir}/r_c$  considering an antenna FWHM= 1 arcmin. In the same plot we show
some upper limits from observations which correspond
to experiments with similar characteristics.

A conclusion can be drawn from Figure~5, the rms fluctuations do not depend
very much on the ratios between the total and the virial radius as well as the
ratio between the virial  and the core radius. However a slight increase occurs
in the  fluctuations when these ratios are smaller since then the gas density
and, therefore, the optical depth are larger.

The other interesting conclusion is that at submillimetre wavelengths
($\lambda<300 \mu$) the SZ effect is quite more important. However, unless we
have grossly overestimated the $f_{gas}$ parameter these SZ fluctuations will
never dominate over primordial temperature fluctuations (typically $\sim
10^{-5}$).

In Figure~6 we show the rms temperature fluctuations as a function of
the  beam FWHM for a wavelength $\lambda= 0.3$~mm, $R=R_{vir}$, and $R_{vir}=5
r_c$. These range from $\sim 2$ to $\sim 7\times 10^{-6}$. We also
show in that figure the temperature variation produced by a single cluster at
$z=0.5$ and different temperatures when the beam is aligned with the cluster
centre. It is clear that for a $\sim 1$ arcminute beam, even a cluster of
$T\sim 1$ keV would be easily detectable since confusion does not operate here
(there are $\sim 3$ clusters per square degree in total). These and any hotter
clusters will then be easily detected and removed from the CMB fluctuation
analysis. If we then recompute the rms fluctuations produced by clusters cooler
than $\sim 1$ keV they are of the order of $10^{-8}$.

The situation might be slightly different for larger beam sizes. At 10 arcmin,
the rms temperature fluctuations are larger and the effect produced by a single
cluster is somewhat diluted. Clusters with temperatures $T\approxlt 3$ keV
might be confused, but those with gas temperatures $T\approxgt 5$ keV would be
easily detectable (a $>10\sigma$ signal at the cluster centre.
Again removing these hotter objects, the rms temperature
fluctuations in the residual CMB do not exceed $\sim 10^{-7}$.

A general conclusion can then be drawn from this study. Clusters of galaxies
produce negligible rms temperature fluctuations in the CMB sky through
Sunyaev--Zel'dovich effect at long wavelengths ($\lambda \approxgt 1$~mm). At
shorter wavelengths, the effect is larger but it never produces anything
equivalent to confusion noise (which could contribute to true background
noise). Instead, the hottest clusters which dominate these fluctuations, appear
as clear signals in sky maps, and will therefore be removed in any CMB
fluctuation analysis. The contribution of the undetected objects to the
residual CMB fluctuations is negligible.

These conclusions could be different if the evolution undergone by clusters is
not well represented by equation (8). In that sense we have evaluated the
changes produced in the rms temperature fluctuations when the parameter $b_2$
(responsible for the evolution with redshift) in $B$ takes different values.
The results obtained are shown in Figure~7. For both beam sizes (FWHM =1~arcmin
and 10~arcmin) the fluctuations increase as the parameter $b_2$ decreases i.e.,
the fluctuations are larger when negative evolution is weaker. Nevertheless
this change is small (a factor of 2-3 at most) and therefore our conclusions
remain unchanged for any reasonable value of the parameter $b_2$.

\medskip

\noindent{\bf 4 DISCUSSION}

\smallskip

\noindent The simple parametrisation of cluster structure and redshift
evolution  used here leads to the conclusion that confusion noise
from the Sunyaev--Zel'dovich effect from these objects does not contribute to
the CMB fluctuations. At the shortest wavelengths under study ($\lambda\sim
0.3$~mm) the hottest clusters which produce the largest fluctuations will be
clearly isolated by any instrument aiming at a detection of CMB angular
structure on arcmin scales. The less massive undetected clusters will produce
negligible background noise. For large beamsizes ($\sim 10$~arcmin), confusion
will be more relevant and only the hottest clusters ($T_{gas}\approxgt 5$~keV)
will be detectable. But even in that case, the remaining undetected objects
will produce negligible sky noise.

Deviations from the picture presented here might arise for various reasons.
First, if the cluster gas has inhomogeneities on top of the smooth profile
assumed here, less Sunyaev--Zel'dovich effect will arise, since gas
inhomogeneities enhance the X--ray luminosity. The same applies to cooling flow
clusters which will produce less Compton scattering than non--cooling flow ones
for a given X--ray luminosity.

The next point to take into account is the effect of clustering of clusters in
our study. Assuming that the correlation function $$\xi(r)=\left(r\over
r_0\right)^{-1.8}\eqno(23)$$ (where the correlation length $r_0\sim 40\,
h_{50}$~Mpc) can be extrapolated down to arbitrarily small separations we can
estimate the enhancement of the CMB temperature fluctuations. If we assume that
the
correlation length corresponds to 80 arcmin at $z=1$ it is easy to estimate
that the probability of finding two clusters in the same beam due to their
clustering is  $\sim 10^{-3}$ for one arcmin beam and $\approxlt  10 $ per cent
for a 10 arcmin beam. That means that on average the clustering can be
completely neglected for beamsizes of a few arcmin and that the corrections
introduced for a beam size of 10 arcmin are quite small.

Our results heavily rely on the cluster properties obtained from the X--ray
data. In particular the flatness of the cluster X--ray source counts implies
that the SZ fluctuations on the CMB are dominated by the X--ray brightest and,
therefore hottest clusters ( Markevitch et al. 1991 also reach the same
conclusions). The situation may be quite different if large
numbers of faint X--ray clusters were present in the sky, i.e., if the cluster
X--ray source counts would raise again at fluxes below $\sim
10^{-11}\ergpcmsqps$. In that case, for which there is no observational support
whatsoever, the SZ
effect on the CMB would be larger and indeed more similar to true confusion
noise.

So far, available X--ray data on clusters tell us that the SZ effect from these
objects both in the spectrum and in the fluctuations of the CMB will not be
relevant at any wavelength.

\bigskip

\noindent {\bf ACKNOWLEDGMENTS}

\noindent We thank L. Cay\'on and E. Mart\'\i nez-Gonz\'alez for discussions
and help. We especially thank the referee, Dr. C. Jones-Forman, for her
enlightening comments which resulted in important improvements of the paper.
Partial financial support for this work was provided by the DGICYT
project PB92-0501 and by the Comission of the EC under contract CHRX-CT92-0033.

\vfill\eject

\noindent {\bf REFERENCES}

\ref Bahcall N.A.,  Soneira R.M., 1983, ApJ, 270, 20

\ref Bartlett J.,  Silk J., 1994, in Sanz J.L., Mart\'\i nez-Gonz\'alez E.,
Cay\'on L., eds., Present and future of the Cosmic Microwave Background,
Springer--Verlag, Heilderberg

\ref Briel U.G., Henry J.P., Schwarz R.A., B\"ohringer H., Ebeling H.,
Edge A.C., Hartner G.D., Schindler S., Tr\"umper J., Voges W., 1991, A\&A,
246, L10

\ref Briel U.G., Henry J.P., B\"ohringer H., 1992,  A\&A, 259, L31

\ref Cole S., Kaiser N., 1988, MNRAS, 233, 637

\ref Davis D.S., Mushotzky R.F., 1993, preprint

\ref David L.P., Slyz A., Jones C., Forman W., Vrtilek S.D., 1993, ApJ,
412, 479

\ref Edge A.C., Stewart G.C., Fabian A.C., Arnaud K.A., 1990, MNRAS, 245,
559

\ref Edge A.C., Stewart G.C., 1991, MNRAS, 252, 414


\ref Evrard A.E., Henry J.P., 1991, ApJ, 383, 95

\ref Fomalont E.B., Kellermann K.I., Anderson M.C., Weistrop D., Wall
J.V., Windhorst R.A., Kristian J.A., 1988, AJ, 96, 1187

\ref Fomalont E.B., Partridge, R.B., Lowenthal, J.D., Windhorst R.A., 1993,
AJ, 404, 8

\ref Forman W., Bechtold J., Blair W., Giacconi R., Van Speybroeck L.,
Jones C. 1981, ApJL, 243, L133.

\ref Gioia I.M. et al., 1982, ApJ, 255, L17

\ref Gioia I., Henry J.P., Maccacaro T., Morris S.L., Stocke J., Wolter
A., 1990a, ApJ, 365, L35

\ref Gioia I., Maccacaro T., Schild R.E., Wolter A., Stocke J., Morris S.L.,
Henry J.P., 1990b, ApJS, 72, 567

\ref Henry J.P., Gioia I.M., Maccacaro T., Morris S.L., Stocke J., Wolter A. ,
 1992, ApJ, 386, 408

\ref Henry J.P. 1992, in Fabian A.C., ed., Clusters and Superclusters
of Galaxies. Kluwer Academic Publishers, Dordrecht, p.311

\ref Jones C., Forman W., 1984, ApJ, 276, 38

\ref Jones C., Forman W., 1992, in Fabian A.C., ed., Clusters and Superclusters
of Galaxies. Kluwer Academic Publishers, Dordrecht, p.49

\ref Kaiser N., 1991, ApJ, 383, 104

\ref Makino N.,  Suto Y., 1993, ApJ, 405, 1

\ref Markevitch M., Blumenthal G.R., Forman W., Jones C., Sunyaev R.A., 1992,
ApJ, 395, 326

\ref Markevitch M., Blumenthal G.R., Forman W., Jones C., Sunyaev R.A., 1991,
ApJ, 378, L33

\ref Mather J.C. et al., 1994, ApJ, 420, 439

\ref Mushotzky R., 1993,  Annals of NY Academy of Sciences, 688, 184

\ref Peebles P.J.E. 1980,  The Large Scale Structure of the Universe,
Princeton Univ. Press, Princeton

\ref Radford S.J.E., 1993, ApJ,  404, L33

\ref Rybicky G.B.,  Lightman A.P., 1979,  Radiative Processes in
Astrophysics, Wiley, New York

\ref Schwarz R.A., Edge A.C., Voges W., B\"ohringer H., Ebeling H.,  Briel
U.G., 1992, A\&A, 256, L11

\ref Subrahmanyan R., Ekers R.D., Sinclair M.,  Silk J., 1993, MNRAS, 263, 416

\ref Sunyaev R.A.,  Zel'dovich Ya.B., 1972, CASP, 4, 173

\ref Sutherland W. 1988, MNRAS, 234, 159

\ref Weinberg S., 1972,  Gravitation and Cosmology,  John Wiley \& Sons,
New York

\ref White S.D.M., Briel U., Henry J.P.,  1993,  MNRAS,  261, L8

\vfill\eject

\noindent {\bf FIGURE CAPTIONS}

\noindent{\bf Figure~1}: X-ray luminosity vs temperature for a sample of 25
clusters.  The hollow points denote the observed values, whilst the filled ones
the computed luminosities according to our model (errors from the uncertainties
in the temperature).  The solid lines correspond to the expected relation for
$R_{vir}/r_c=5$ (below) and $R_{vir}/r_c=10$ (above) for $f_{gas}=0.15$. Same
values for $f_{gas}=0.06$ are shown as dotted lines.

\medskip

\noindent{\bf Figure~2.}: Computed gas mass within 3 Mpc for a sample of 25
clusters using the measured temperatures, $R=R_{vir}$ and $f_{gas}=0.15$ as a
function of observed gas mass within 3 Mpc (by Jones \& Forman 1984).

\medskip

\noindent{\bf Figure~3}: Temperature variation for 4 different cluster
temperatures: $2.8$~keV (solid), $4.7$~keV (dashed), $6.7$~keV (dot-dashed),
$8.6$~keV (dotted), smeared out with a beam of 1~arcmin FWHM pointing to the
cluster centre. A zero-crossing occurs whenever a curve hits the {\it x}-axis
in this plot. The redshift is $z=0.5$, $R=R_{vir}$ and $R_{vir}=5 r_c$ (see
text for details).

\medskip

\noindent{\bf Figure~4}: Average Compton $<y>$ parameter versus $R/R_{vir}$ for
different $R_{vir}/r_c$ ratios. Solid line at the top is COBE FIRAS upper
limit.

\medskip

\noindent{\bf Figure~5}: The rms fluctuations for a map with FWHM= 1 arcmin.
Solid lines correspond to $R/R_{vir}=1$ and $R_{vir}/r_c=5,10$ (from top to
bottom). Dotted lines are for $R/R_{vir}=2$ and the same for $R_{vir}/r_c$
ratios. The vertical arrows are upper limits from observations with VLA
(Fomalont et al. 1988, with 60~arcsec FWHM, and Fomalont et al. 1993, with a
resolution of 80~arcsec), ATCA (Subrahmanyan, Ekers, Sinclair \& Silk 1993,
with
2.1~arcmin FWHM) and IRAM (Radford 1993, with 55~arcsec FWHM), from right to
left. Due to the different beamsizes of  each experiment only the upper
limit from Fomalont et al. (1988) can be directly compared to the lines plotted
here.

\medskip

\noindent{\bf Figure~6}: The rms temperature fluctuations as a function of the
beam FWHM for a wavelength $\lambda= 0.3$~mm, $R=R_{vir}$, and $R_{vir}=5 r_c$
(individual points). The lines correspond to the temperature variation produced
by a single cluster: $2.8$~keV (solid), $4.7$~keV (dashed), $6.7$~keV
(dot-dashed), $8.6$~keV (dotted).  Asterisks correspond to the
fluctuations produced by objects with $T<1$~keV at a FWHM of 1~arcmin
and objects with $T<5$~keV at a FWHM of 10~arcmin.

\medskip

\noindent{\bf Figure~7}: The rms temperature fluctuations as a function of the
parameter $b_2$ (see text for details) for beams of 1~arcmin FWHM (solid line)
and 10~arcmin FWHM (dashed line).
\bye
\end